\documentstyle[aps,amstex,amssymb,graphics,preprint,epsf]{revtex}
\tightenlines
\begin{document}                                                       

\draft 


\title {The Euclidean resonance and quantum tunneling}

\author{B.I.\ Ivlev}

\address{Department of Physics and Astronomy\\
University of South Carolina, Columbia, SC 29208\\
and\\
Instituto de F\'{\i}sica, Universidad Aut\'onoma de San Luis Potos\'{\i}\\
San Luis Potos\'{\i}, S. L. P. 78000 Mexico}


\maketitle

\begin{abstract}
The extremely small probability of tunneling through an almost classical potential barrier may become not small under the
action of the specially adapted non-stationary signal which selects the certain particle energy $E_{R}$. For particle
energies close to this value, the tunneling rate is not small during a finite interval of time and has a very sharp 
peak at the energy $E_{R}$. After entering inside the barrier, the particle emits electromagnetic quanta and exits the 
barrier with a lower energy. The signal amplitude can be much less compared to the field of the static barrier. This 
phenomenon can be called the Euclidean resonance since the under-barrier motion occurs in imaginary time. The resonance 
may stimulate chemical and biochemical reactions in a selective way by adapting the signal to a certain particular 
chemical bond. The resonance may be used in search of the soft alpha-decay for which a conventional observation is 
impossible due to an extremely small decay rate.

\end{abstract}
\vspace{0.7cm} 


\narrowtext
\section{INTRODUCTION}
Quantum tunneling under a potential barrier is a part of a variety of physical processes. The computation of the 
probability for a classically forbidden region has a certain peculiarity from the mathematical standpoint: there 
necessarily arises here the concept of motion in imaginary time or along a complex trajectory 
\cite{LANDQUANT,POKROV,COLEMAN}. The famous semiclassical approach of Wentzel, Kramers, and Brillouin (WKB) \cite{LANDQUANT}
for tunneling probability can be easily reformulated in terms of classical trajectories in the complex time. The method of
complex trajectories is also applicable to a non-stationary case \cite{KELD,PERELOM}. This method has been further 
developed in papers \cite{MEL1,MEL2,MEL3,MEL4,MEL5}, where singularities of the trajectories in the complex plane were
accounted for an arbitrary potential barrier (see also \cite{MILLER}). Recent achievements in the semiclassical theory are
presented in Refs.~\cite{KESHA,DEFENDI,MAITRA,ANKER,CUNIBER}.

A control of tunneling processes by external signal is a part of the field called quantum control which is actively 
developed now, see for example Ref.~\cite{RAB1} and references therein. Excitations of molecules, when one should excite
only particular chemical bonds \cite{RAB2,KRAUSE1}, a control of acoustic waves in solids \cite{RAB3}, formation of 
programmable atomic wave packets \cite{SCHU}, a control of electron states in heterostructures \cite{KRAUSE2}, and a control
of photo current in semiconductors \cite{ATAN}, are typical examples of control by laser pulses.

Let us focus on main aspects of tunneling under nonstationary conditions. When the electric field ${\cal E}\cos\Omega t$ acts
on a tunneling particle of the initial energy $E$, it can absorb the quantum $\Omega$ (with the probability proportional to
the small parameter ${\cal E}^{2}$) and tunnel after that in the more transparent part of the barrier with the higher 
energy $E+\Omega$. The pay in the absorption probability may be compensated by the probability gain in tunneling. In this 
case the system tends to absorb further quanta to low the total probability of barrier penetration. If $\Omega$ is not 
big, the process of tunneling, with a simultaneous multi-quanta absorption, can be described in a semiclassical way by the
method of classical trajectories in the complex time \cite{MEL1,MEL2,MEL3,MEL4,MEL5}. When a tunneling particle of the 
energy $E$ is acted by a short-time signal, the particle energy after escape is $E+\delta E$, where the energy gain 
$\delta E=N\omega$, should be extremized with respect to the number of absorbed quanta $N$ and the energy $\omega$ of each 
quantum. This extreme corresponds to the method of classical trajectories \cite{ADAPT}.

According to the perturbation theory, the both probabilities, of absorption and emission of a quantum, are small being 
proportional to ${\cal E}^{2}$. After absorption, $\delta E$ is positive which enhances the total probability due to the 
increase of the tunneling rate. After emission, $\delta E$ is negative and the particle should tunnel with a smaller energy
in a less transparent part of the barrier; in this case there is no gain in the probability due to tunneling as for 
absorption. At first sight, tunneling in a nonstationary field cannot be assisted by an emission of quanta of this field 
since one should pay in probability twice. This conclusion is correct as soon as the nonstationary field is small and the 
perturbation theory is applicable. For tunneling the perturbation theory with respect to a non-stationary field stops to 
be valid already for very small fields \cite{MEL1,MEL2,MEL3,MEL4,MEL5,ADAPT}. What happens to emission-assisted tunneling 
when the nonstationary field is not very small?

For a not very small nonstationary field the perturbation approach should be substituted by the semiclassical one. In
this approach a multi-quanta absorption is accounted by classical trajectories in the complex time and the transition
probability $W\sim\exp(-A)$ is expressed through the classical action $A$ calculated by means of those trajectories. This 
differs from the the stationary case described by the WKB formula $W\sim\exp[-A_{0}(E)]$. The particle with the initial energy
$E$ emits quanta, ``dives'' on the level $E-\mid\delta E\mid$, and proceeds the penetration with this energy during the 
certain time interval $i\theta$. The ``diving'' process does not contribute much to the action $A$ (in contrast to the 
perturbation theory) but the further motion adds $A_{0}(E-\mid\delta E\mid)$ and the term $2\theta\delta E <0$. The last term is 
analogous to the occupation one, $\delta E/T$, when the temperature $T$ is substituted by $1/2\theta$ \cite{CALDLEGG}. 
As shown in the paper, under some conditions of the nonstationary signal the ``diving'' process is instant in the complex 
time and weakly contributes to the total action. The condition of the instant ``diving'' is not necessary for the strong 
reduction of $A$. 

The emission-assisted tunneling exhibits a surprising feature: under action of the certain signal the tunneling rate can be
not exponentially small since the action can tend to zero ($A(E_{R})=0$) at some energy $E_{R}$ which depends on the
signal parameters. The negative term $2\theta\delta E$ in the action plays an important role in the dramatic reduction of the
action $A$ under certain conditions. According to this, the tunneling rate $W\sim\exp(-A)$ reaches a sharp maximum at the 
energy $E_{R}$. The method used implies formally a small probability $\exp(-A)\ll 1$ but it can be of the order of $10^{-1}$
instead of an extremely small value. In an anharmonic potential well a classical or quantum particle of the energy $E$, 
acted by the signal ${\cal E}\cos\Omega t$, can exhibit a resonance if to adapt the frequency $\Omega$ to the energy $E$ 
\cite{LANDMECH,LANDQUANT}. Analogously, to get a maximum tunnel penetration at some given initial energy $E$ of the 
particle, the external signal should be adapted to this energy. In the first case a motion occurs in real time but in 
the second case one should use imaginary (complex) time to describe the process. Traditionally, an action in imaginary 
time is called the Euclidean action which comes from relativity where the geometry is Euclidean in imaginary time. 
Extending this analogy, one can call the above tunnel phenomenon the Euclidean resonance.

A scenario of the Euclidean resonance is the following: the particle with the energy $E$, before the barrier, leaves it 
having the exit energy $E+\delta E$ which is less than $E$; the position of the exit wave packet in time is close to the 
peak of the signal and the duration of the exit wave packet is roughly the under-barrier time $\theta$; during this time 
interval the tunneling rate is not exponentially small for energies $(E_{R}-E)/E_{R}\sim A^{-1}_{0}(E_{R})$. This means that under 
action of a specially adapted signals a particle can penetrate through very non-transparent barriers (big $A_{0})$. One 
should emphasize, this process corresponds to a motion deep under a barrier but not a transition over its top. As shown in
the paper, the signal amplitude, required for the Euclidean resonance, can be very small comparing to an electric field of
the static barrier. 

In Secs. II-IV the problem of the Euclidean resonance for a decay of the metastable state in the $\delta$-potential 
well is solved exactly in the semiclassical sense. This means that for the wave function 
$\psi(x,t)=\exp[iS(x,t)/\hbar+i\sigma_{1}(x,t)+i\hbar\sigma_{2}(x,t)+...]$ (we use $\hbar$ in this relation) the action 
$S$, the first, and the second non-semiclassical correction $\sigma_{1}$ and $\sigma_{2}$ have been found exactly in the 
analytical forms. The developed method enables to write down the exact expression for any $\sigma_{n}$. In Sec. V the 
semiclassical condition $(\sigma_{1}\gg\sigma_{2}\gg ...)$ is obtained. In Sec. VI the method of complex trajectories is 
shown to be valid for the problem. In Secs. VIII-IX the method of complex trajectories is applied to the Euclidean 
resonance in a smooth potential. In Sec. XI some particular manifestations of the resonance are considered. It can be used
for a selective stimulation of chemical reactions. For alpha-decay of nuclei the experimentally registered energies of 
alpha-particles range in the approximate interval of 2-10 MeV. For lower energies the tunneling probability is very low 
which makes observations to be impossible. However, there is no a principal ban for the soft alpha-decay, for example, 
with the energies of alpha-particles of the order of 1 KeV. The energy, counted from the bottom of the nuclear potential 
well, remains of the order of MeV. Adapting an external signal, with the duration of one or ten femtoseconds, to a 
possible energy level of an alpha-particle, one can try to search the soft alpha-decay. In principle, one can stimulate 
the conventional alpha-decay but this requires a very short signals of the order of $10^{-19}{\rm s}$.
\section{TRIANGULAR BARRIER}
As in Ref.~\cite{ADAPT}, we consider decay of the metastable state in the potential
\begin{equation}
\label{1}
V(x)=V-{\cal E}_{0}\mid x\mid -\sqrt{\frac{2(V-E)}{m}}\hspace{0.1cm}\delta(x)
\end{equation}
under action of a nonstationary electric field ${\cal E}(t)$. In the limit ${\cal E}_{0}\rightarrow 0$, the energy $E$ 
corresponds to the bound state in the $\delta$-function potential well. The symmetric wave function 
$\psi(x,t)=\exp[iS(x,t)+i\sigma(x,t)]$ depends on the classical action $S$ and the quantum correction $\sigma$. The 
action $S$ has the form \cite{ADAPT}
\begin{align}
\label{2}
S(x,t)=&\frac{{\cal E}^{2}_{0}}{2m}\int^{t_{0}}_{t}dt_{1}\left[i\tau_{00}-(t_{0}-t_{1})-
\int^{t_{0}}_{t_{1}}dt_{2}h(t_{2})\right]^{2}\nonumber\\
&+{\cal E}_{0}x\left[i\tau_{00}-(t_{0}-t)-\int^{t_{0}}_{t}dt_{1}h(t_{1})\right]+(V-E)t-Vt
\end{align}
where
\begin{equation}
\label{3}
\tau_{00}=\frac{\sqrt{2m(V-E)}}{{\cal E}_{0}}\hspace{0.1cm};\hspace{2cm}h(t)=\frac{{\cal E}(t)}{{\cal E}_{0}}
\end{equation}
and the function $t(x,t)$ is defined by the equation
\begin{equation}
\label{4}
\frac{mx}{{\cal E}_{0}}=i(t-t_{0})\tau_{00}+\frac{(t_{0}-t)^{2}}{2}+\int^{t_{0}}_{t}dt_{1}(t_{1}-t)h(t_{1})
\end{equation}
For a symmetric pulse ${\cal E}(-t)={\cal E}(t)$ the moment $t=0$ plays an important role and it is convenient to 
introduce $\tau_{0}(x)=-it_{0}(x,0)$. According to Eq.~(\ref{4}), the function $\tau_{0}(x)$ is determined by the relation
\begin{equation}
\label{5}
x=\frac{{\cal E}_{0}}{2m}\left[2\tau_{0}\tau_{00}-\tau^{2}_{0}-2\tau_{00}\int^{\tau_{0}}_{0}d\xi h(\xi)\right]
\end{equation}
By means of Eqs.~(\ref{2}) and (\ref{5}) one can write down the first derivative of the action
\begin{equation}
\label{6}
i\frac{\partial S(x,0)}{\partial x}={\cal E}_{0}\left[\varphi (\tau_{0})-\tau_{00}\right]
\end{equation}
and the second derivative
\begin{equation}
\label{7}
i\frac{\partial^{2}S(x,0)}{\partial x^{2}}=m\hspace{0.1cm}\frac{1+h(\tau_{0})}{\tau_{00}-\tau_{0}-\tau_{0}h(\tau_{0})}
\end{equation}
Here $\tau_{0}$ is expressed through $x$ by Eq.~(\ref{5}) and
\begin{equation}
\label{8}
\varphi (\tau)=\tau +\int^{\tau}_{0}d\xi h(\xi)
\end{equation}
In the static case, ${\cal E}(t)=0$, the action has the WKB form \cite{LANDQUANT}
\begin{equation}
\label{9}
iS=\frac{2\tau_{00}}{3}(V-E)\left[\left(1-\frac{x{\cal E}_{0}}{V-E}\right)^{3/2}-1\right]
\end{equation}
The point $x=x_{1}$, where $i\partial S/\partial x=0$, is remarkable since at this point the particle comes out of the 
barrier. The exit point $x_{1}$ is given by Eq.~(\ref{5}) with $\tau_{0}=\tau_{1}$  where $\tau_{1}$ is determined by the 
equation
\begin{equation}
\label{10}
\varphi (\tau_{0})=\tau_{00}
\end{equation}
There is another remarkable point $x_{2}=x(\tau_{2})$, where $\tau_{2}$ can be found from the equation
\begin{equation}
\label{11}
\tau_{00}-\tau_{0} -\tau_{0}h(\tau_{0})=0
\end{equation}
The first derivative (\ref{6}) is finite at this point but $i\partial^{2}S/\partial x^{2}\sim (x_{2}-x)^{-1/2}$.
\section{THE PROBABILITY OF DECAY}
One can show, using Eq.~(\ref{2}), that
\begin{equation}
\label{12}
iS(x_{1},t)=iS(x_{1},0)+\frac{t^{4}{\cal E}^{2}_{0}}{8m^{2}}\left[i\frac{\partial^{2}S(x,0)}{\partial x^{2}}\right]_{x_{1}}
\end{equation}
Eq.~(\ref{12}) determines the duration of the exit wave packet
\begin{equation}
\label{13}
\delta t=\left(\frac{{\cal E}^{2}_{0}}{8m^{2}}\left[i\frac{\partial^{2}S(x,0)}{\partial x^{2}}\right]_{x_{1}}\right)^{-1/4}
\end{equation}
Only those exit points $x_{1}$ are physical for which the second derivative of the action is negative. The classical 
trajectory is given by the Newton equation 
\begin{equation}
\label{14}
m\hspace{0.1cm}\frac{\partial^{2}x(t)}{\partial t^{2}}={\cal E}_{0}+{\cal E}(t)
\end{equation}
One can show from the Hamilton-Jacoby equation \cite{LANDMECH}, that
\begin{equation}
\label{15}
{\rm Im}\hspace{0.1cm}\frac{dS(x(t),t)}{dt}=
\frac{m}{2}\hspace{0.1cm}{\rm Im}\left(\frac{\partial x}{\partial t}\right)^{2}+
\left[{\cal E}_{0}+{\cal E}(t)\right]{\rm Im}\hspace{0.1cm}x(t)
\end{equation}
The qualitative plot of ${\rm Im}S$ is shown in Fig.~1. We see that the imaginary part of the action is a constant at a 
real classical trajectory. One can find the probability $\int dx\mid\psi\mid^{2}$ to find the particle outside the 
barrier, after an action of the pulse, by putting $\psi =\exp(iS+i\sigma)$, where $\sigma $ is a non-semiclassical part in
the exact expression of the wave function \cite{ADAPT}. At every moment of time the $x$-integration is determined by the 
saddle point in the vicinity of the classical trajectory in Fig.~1 and does not depend on time, according to conservation 
of flux. For this reason, one can calculate the probability at $t=0$ when
\begin{equation}
\label{16}
\psi=\exp\left[iS(x_{1},0)+\frac{(x-x_{1})^{2}}{2}\left(i\frac{\partial^{2}S}{\partial x^{2}}\right)_{x_{1}}+i\sigma_{1}(x_{1},0)\right]
\end{equation}
According to Eq.~(23) of Ref.~\cite{ADAPT}, the first non-semiclassical correction has the form
\begin{equation}
\label{17}
i\sigma_{1}(x_{1},0)=-\ln\sqrt{1-\frac{\tau_{1}}{\tau_{00}}\left[1+h(\tau_{1})\right]}-\frac{1}{2}
\end{equation}
Ignoring the finite duration $\delta t$ (\ref{13}) of the exit wave packet and using Eq.~(\ref{7}) for the second 
derivative, one can obtain for the total probability to find the particle outside the barrier the following expression
\begin{align}
\label{18}
&W(E,t)=t(V-E)\exp(-A_{0})\nonumber\\
&+\frac{4}{e}(V-E)\frac{\sqrt{2\pi m}}{{\cal E}_{0}}\hspace{0.1cm}\frac{\exp\left[-A(E)\right]}{\sqrt{(\tau_{1}h(\tau_{1})+\tau_{1}
-\tau_{00})(1+h(\tau_{1}))}}\hspace{0.1cm}\Theta\left[t-\sqrt{\frac{2m}{{\cal E}_{0}}(x_{det}-x_{1})}\hspace{0.1cm}\right]
\end{align}
where $\Theta(t)$ is the step function and
\begin{equation}
\label{19}
A(E)=2\hspace{0.1cm}{\rm Im}S(x_{1},0)
\end{equation}
The first term corresponds to the conventional decay process $({\cal E}(t)=0)$ \cite{LANDQUANT}, when
\begin{equation}
\label{20}
A_{0}=\frac{4}{3}(V-E)\tau_{00},
\end{equation}
It comes from the continuity equation 
\begin{equation}
\label{21}
\frac{\partial W}{\partial t}=-\frac{2}{m}\hspace{0.1cm}{\rm Im}\hspace{0.1cm}\psi^{*}\frac{\partial\psi}{\partial x}\hspace{0.1cm};
\hspace{1cm}\psi=\frac{\sqrt{m(V-E)}}{p(x)}\exp\left(i\int pdx\right)
\end{equation}
where the momentum has its classical form $p(x)=\sqrt{2m(E-V+{\cal E}_{0}x)}$. In Eq.~(\ref{18}) $x_{det}$ is the point of 
detection of the exit particle. Outside the barrier a correction of the trajectory (\ref{14}) by the non-stationary part 
${\cal E}(t)$ is small. The first term in Eq.~(\ref{17}) is a short-time expansion of the full expression 
$1-\exp[-t(V-E)\exp(-A_{0})]$ corresponding to the multi-instanton approach.
\section{THE NONSTATIONARY PULSE}
Now let us specify a particular shape of the pulse of the electric field
\begin{equation}
\label{22}
{\cal E}(t)=-{\cal E}\exp\left(-\Omega^{4}t^{4}-2\Omega^{4}\theta^{2}t^{2}\right)
\end{equation}
We consider further the limit $\Omega^{4}\theta^{4}\gg 1$ and, omitting the small width 
$(\Omega^{2}\theta)^{-1}\ll\theta$, one can write
\begin{equation}
\label{23}
\varphi (\tau)=\tau-\lambda\theta\Theta(\tau -\theta);\hspace{1cm}
\lambda =\frac{\sqrt{\pi}}{2\Omega^{2}\theta^{2}}\hspace{0.1cm}\frac{{\cal E}}{{\cal E}_{0}}\exp\left(\Omega^{4}\theta^{4}\right)
\end{equation}
For the pulse (\ref{22}) the function $\varphi^{2}(\tau)$ is plotted in Fig.~2. One can see that at some values of 
$\tau^{2}_{00}$ three solutions of the equation $\varphi^{2}(\tau)=\tau^{2}_{00}$ are possible. According to this, there are 
different types of $x$-dependence of $iS(x,0)$ which are shown in Fig.~3. The thick solid curves in Fig.~2 represents the 
physical regime when the second derivative (\ref{7}) is negative. The regimes, corresponding to the thin solid curves 
in Fig.~2, are not realized. This refers only to the vicinity of the moment $t=0$; when a moment of time is not close to 
zero, the pulse is small and there is only one physical branch $\varphi^{2}(\tau)=\tau^{2}$. We do not study here how the two 
regimes merge with time and consider only a time interval bigger than $\delta t$ when the equation (\ref{17}) for the 
probability holds.
\section{THE SEMICLASSICAL CONDITION}
The semiclassical condition means a big action $S$ comparing to the non-semiclassical part $\sigma$ \cite{LANDQUANT}. As
follows from the expression for $\sigma$ \cite{ADAPT}, it becomes big at the point $x_{2}$ where the second derivative of 
the action (\ref{7}) is infinity. The first derivative (\ref{6}) remains finite at this point which makes an essential 
difference with the static case (${\cal E}(t)=0$) when the relations $i\partial S/\partial x=0$ and 
$i\partial^{2}S/\partial x^{2}=\infty$ are fulfilled at the same point giving rise to the Stokes lines \cite{HEADING}. In 
the present case there is no Stokes lines which makes the procedure of going around $x_{2}$ in the complex plane to be 
easy. This procedure enables to avoid the point $x_{2}$, where the semiclassical condition breaks down, and to go over from
one branch in Fig.~3 to another remaining within the semiclassical approach. For the pulse (\ref{22}), the semiclassical 
condition does not break down at the point $x_{1}$.

For the pulse (\ref{22}), Eqs.~(\ref{10}) and (\ref{21}) result in the estimates
\begin{equation}
\label{24}
\tau_{1}-\theta\sim\frac{1}{\Omega^{2}\theta}\hspace{0.1cm};\hspace{1cm}\tau_{2}-
\theta\simeq\frac{\sqrt{\ln\Omega^{2}\theta^{2}}}{2\Omega^{2}\theta}
\end{equation}
Going around the point $x_{2}$, one can put $\tau_{0}=\tau_{2}+\delta\tau$, where 
$\delta\tau =\mid\delta\tau\mid\exp(i\phi)$ and the phase $\phi$ varies between $0$ and $\pi$. We have
\begin{equation}
\label{25}
\tau_{00}-\tau_{0}-\tau_{0}h(\tau_{0})\simeq
\left(\tau_{00}-\theta\right)\left[1-\exp\left(-4\Omega^{2}\theta\sqrt{\ln\Omega^{2}\theta^{2}}\hspace{0.1cm}\delta\tau\right)\right]
\end{equation}
One can choose $\mid\delta\tau\mid\sim(\Omega^{2}\theta)^{-1}$. In the semiclassical expansion 
$\sigma =\sigma_{1}+\sigma_{2}+...$ it should be $A\gg\sigma_{1}\gg\sigma_{2}\gg ...$ The estimations at the semi circle 
around $\tau_{2}$, easily followed from \cite{ADAPT}, are
\begin{equation}
\label{26}
i\sigma(\tau_{2}+\delta\tau,0)\sim\frac{1}{2}\ln\frac{\tau_{00}}{\tau_{00}-\theta}-\frac{\theta}{2\tau_{00}}\hspace{0.1cm}(1-\lambda)
\end{equation}
and
\begin{equation}
\label{27}
i\sigma(\tau_{2}+\delta\tau)\sim
\frac{\tau_{00}}{(V-E)\theta}\hspace{0.1cm}\frac{\Omega^{4}\theta^{4}\ln\Omega^{2}\theta^{2}}{\tau_{00}-\theta}
\end{equation}
Now the semiclassical condition $\sigma_{1}\gg\sigma_{2}$ reads
\begin{equation}
\label{28}
\frac{\tau_{00}}{\theta -\tau_{00}}\hspace{0.1cm}\Omega^{4}\theta^{4}\ln\Omega^{2}\theta^{2}\ll(V-E)\theta\hspace{0.1cm}
\end{equation}
The method used implies that $\exp(-A)\ll 1$ but this parameter can be of the order of $10^{-1}$ instead of an extremely 
small value.
\section{THE METHOD OF COMPLEX TRAJECTORIES}
The action $A$ in Eq.~(\ref{17}) can be written in the form
\begin{equation}
\label{29}
A=-2i\int^{\tau_{1}}_{0}d\tau_{0}\hspace{0.1cm}\frac{\partial S(x(\tau_{0}),0)}{\partial \tau_{0}}
\end{equation}
where $x(\tau_{0})$ is given by Eq.~(\ref{5}). The relation (\ref{29}) can be transformed into the following one
\begin{equation}
\label{30}
A= \frac{{\cal E}^{2}_{0}}{m}\left[\tau_{1}{\varphi (\tau_{1})}^{2}-\int^{\tau_{1}}_{0}d\tau_{0}{\varphi (\tau_{0})}^{2}\right]
\end{equation}
where $\tau_{1}$ is determined by Eq.~(\ref{10}).

The result (\ref{30}) can be obtained by the method of trajectories in the imaginary (corresponding to the Euclidean norm)
time $t=i\tau$
\begin{equation}
\label{31}
A=2\int^{\tau_{1}}_{0}d\tau \left[\frac{m}{2}\left(\frac{\partial x}{\partial\tau}\right)^{2}+V(x)-{\cal E}(t)x-E\right]
\end{equation}
where
\begin{equation}
\label{32}
V(x)=V-{\cal E}_{0}x
\end{equation}
The trajectory satisfies the classical equation in the imaginary (Euclidean) time
\begin{equation}
\label{33}
m\hspace{0.1cm}\frac{\partial^{2}x(\tau)}{\partial\tau^{2}}=-{\cal E}_{0}-{\cal E}(\tau)
\end{equation}
with the boundary conditions
\begin{equation}
\label{34}
\left(\frac{\partial x(\tau)}{\partial\tau}\right)_{\tau_{1}}=-\sqrt{\frac{2(V-E)}{m}}\hspace{0.1cm};\hspace{1cm}
x(\tau_{1})=0\hspace{0.1cm};\hspace{1cm}\left(\frac{\partial x(\tau)}{\partial\tau}\right)_{0}=0
\end{equation}
It follows from Eqs.~(\ref{33}) and (\ref{34}) that
\begin{equation}
m\hspace{0.1cm}\frac{\partial x}{\partial\tau}=-{\cal E}_{0}-\int^{\tau}_{0}d\xi {\cal E}(\xi)
\end{equation}
\label{35}
The trajectory has the form
\begin{equation}
\label{36}
x(\tau)=\frac{{\cal E}_{0}}{2m}\left[\tau^{2}_{1}-\tau^{2}+2\int^{\tau_{1}}_{0}d\xi\int^{\xi}_{0}d\eta h(\eta)\right]
\end{equation}
The trajectory starts at the exit point $x(0)=x_{1}$, with the energy $E+\delta E$, and terminates at the enter point 
$x(\tau_{1})=0$ with the energy $E$. The action (\ref{31}), in the method of classical trajectories, produces the same 
action (\ref{30}) followed from the Hamilton-Jacobi equation. The method of classical trajectories should be supplemented 
by the rule of selection of stable extrema
\begin{equation}
\label{37}
i\left(\frac{\partial^{2}S(x,0)}{\partial x^{2}}\right)_{x_{1}}<0
\end{equation}
The energy of the exit particle $E+\delta E$ can be found from the condition
\begin{equation}
\label{38}
E+\delta E=V-{\cal E}_{0}x_{1}
\end{equation}
and the energy gain is
\begin{equation}
\label{39}
\delta E=\frac{(V-E)}{\tau_{00}}\left[\tau^{2}_{00}-\tau^{2}_{1}-2\int^{\tau_{1}}_{0}d\tau_{0}\int^{\tau_{0}}_{0}d\xi h(\xi)\right]
\end{equation}
\section{THE EUCLIDEAN RESONANCE}
The action (\ref{30}) has the geometric interpretation
\begin{equation}
\label{40}
A=\frac{{\cal E}^{2}_{0}}{m}\left(Q_{+}-Q_{-}\right)
\end{equation}
where $Q_{+}$ and $Q_{-}$ are the areas shown in Fig.~2 for the pulse (\ref{22}). The plot of the action (\ref{40}) is shown
in Fig.~4 where the thick curves correspond to the physical regimes under the condition (\ref{37}). The action satisfies 
the relation $\partial A/\partial E =\partial A_{0}/\partial E$, at $\tau_{00}=\theta$, where $A_{0}$ is given by 
Eq.~(\ref{20}). The action $A$ has to be inserted in Eq.~(\ref{18}) and at a given $\tau_{00}$ one should take the lower 
curve. In this sense the regions (b) and (c) in Fig.~4 are most important where
\begin{equation} 
\label{41}
A=2(E_{R}-E)\theta\hspace{0.1cm};\hspace{1cm}E<E_{ext}
\end{equation}
Here the definitions are used
\begin{equation}
\label{42}
E_{ext}=E_{R}-\frac{\theta^{2}{\cal E}^{2}_{0}}{6m}\left(1+\frac{1}{\sqrt 3}-
\lambda\right)\left(\lambda_{T}-\lambda\right)\Theta\left(\lambda_{T}-\lambda\right)
\end{equation}
where the resonance energy and the threshold pulse amplitude are introduced
\begin{equation}
\label{43}
E_{R}=V-\frac{\theta^{2}{\cal E}^{2}_{0}}{6m}\hspace{0.1cm};\hspace{1cm}\lambda_{T}=1-\frac{1}{\sqrt 3}
\end{equation}
Within the exponential accuracy the probability (\ref{18}) to find the particle outside the barrier is
\begin{equation}
\label{44}
W\sim\exp\left[-2(E_{R}-E)\theta\right]\Theta\left(E_{ext}-E\right)
\end{equation}
As shown in Fig.~4, the expression (\ref{44}) has the jump behavior as a function of $E$. At $E_{ext}<E$ the exit point 
$x_{1}$ becomes complex resulting in the complex trajectory, defined by Eq.~(\ref{14}) in the real time, after exit the 
barrier. As follows from Eq.~(\ref{15}), the amplitude of the exit wave packet exponentially decays in time as for a 
particle with a complex energy. Hence the particle with $E>E_{ext}$ does not leave the barrier.

The energy of the exit particle (for the lower curve in Fig.~4) 
\begin{equation}
\label{45}
E+\delta E=V-3(V-E_{R})
\end{equation}
turns to $E$ at the upper end of the lower curve in Fig.~4. The duration of the exit wave packet (\ref{13}) in the case of 
the pulse (\ref{22}) is
\begin{equation}
\label{46}
\delta t=\theta\left[\frac{4}{3(V-E_{R})\theta}\right]^{1/4}
\end{equation}
Within the exponential accuracy the tunneling probability 
$W\sim\exp\left(-{\rm min}\left\{A_{0}\hspace{0.1cm};A\right\}\right)$ (\ref{18}) is plotted as a function of energy in 
Fig.~5. At each amplitude of the pulse $\lambda$ there is a peak of $W$ at the energy $E_{ext}$ (\ref{42}). When 
$\lambda\rightarrow\lambda_{T}$ the peak grows, reaches the value $W\sim 1$, and remains of the order of unity under 
further increase of the amplitude $\lambda\geq\lambda_{T}$ as follows from Eq.~(\ref{44}). The function $W(E)$ has a very 
sharp peak at $E=E_{R}$ which jumps down to $W\sim\exp(-A_{0})$ at $E>E_{R}$. Formally, the action $A$ becomes of the 
order of unity at $E=E_{R}$ and the semiclassical condition violates. This means that the energy $E_{R}-E$ should not be 
less than ${\theta}^{-1}$. To get a classic resonance at some fixed energy in a non-harmonic potential one should adapt the 
frequency of the periodic signal to this energy and the oscillation amplitude depends on the strength of the signal. 
Analogously, to get a maximum tunnel penetration at some energy $E_{R}$ the external signal should be also adapted to this 
energy (by means of Eq.~(\ref{43})) and the intensity of the exit particles depends on the signal amplitude. In the first 
case a motion occurs in the real time but in the second case one should use the imaginary (complex) time to describe the 
process. Traditionally, an action in the imaginary time is called the Euclidean action which comes from the relativity 
where the geometry is Euclidean in the imaginary time. Extending this analogy, one can call the above tunnel phenomenon 
the Euclidean resonance.

The non-stationary field (\ref{22}) at $\Omega^{4}\theta^{4}\gg 1$ can be called the instant signal since in the imaginary 
time it is well localized near the ``moment'' $i\theta$. The Euclidean resonance can occur not only for instant signals, 
nevertheless, they have an advantage of enhancement of the effective amplitude 
${\cal E}\exp(\Omega^{4}\theta^{4})\gg {\cal E}$. For an instant signal, if the initial energy level $E$ is fixed, the Euclidean 
resonance occurs at some resonance value $\theta_{R}$ defined by the condition (\ref{43})
\begin{equation}
\label{47}
\theta_{R}=\frac{6m(V-E)}{{\cal E}_{0}}
\end{equation}
The effective amplitude of the signal $\lambda$ (\ref{23}) should satisfy not a resonance but the threshold condition
$\lambda >\lambda_{T}$. The probability of tunneling (\ref{44}) can be rewritten in the form
\begin{equation}
\label{48}
W\sim\exp\left[-3\sqrt{3}A_{0}(E)\frac{\theta_{R}-\theta}{\theta_{R}}\right]\Theta\left(\theta_{R}-\theta\right)
\end{equation}
where $A_{0}(E)$ is the conventional action (\ref{20}).

The way of the particle under the barrier is shown in Fig.~6 in terms of its total energy. In the limit of big 
$\Omega^{4}\theta^{4}$ the right-hand side of Eq.~(\ref{33}) has the form 
$-{\cal E}_{0}+\lambda {\cal E}_{0}\delta (\tau -\theta)$. According to this, the particle moves in the following way: (1)
it starts at the point $x=x_{1}$ ($\tau =0$) and moves free, (2) it reaches the point $x\simeq 0$ approximately at the 
moment $i\theta$, (3) it gets the instant energy gain at the point $x\simeq 0$ at the moment $i\theta$, and (4) it arrives
in the true point $x=0$ having the energy $E$. Since the non-stationary pulse is instant, the action (\ref{31}) does not 
change during the pulse and can be written as a sum of two parts
\begin{equation}
\label{49}
A=A_{0}(E+\delta E)+2\theta\delta E
\end{equation}
where the conventional action $A_{0}$ is given by Eq.~(\ref{20}) under the condition $\tau_{00}=\theta$. Since $\delta E<0$, the 
action increases $A_{0}(E+\delta E)>A_{0}(E)$, but the term $-2\theta\mid\delta E\mid$ provides a reduction of the total action which
results in $A=0$ at $E=E_{R}$.
\section{THE EUCLIDEAN RESONANCE FOR A SMOOTH POTENTIAL}
Suppose the static potential to have no singularities, like the potential (\ref{1}). Such a potential is shown in Fig.~7(a).
In this case the action $A$ has the form
\begin{equation}
\label{50}
A=-2\hspace{0.1cm}{\rm Im}\int_{C}dt\left[\frac{m}{2}\left(\frac{\partial x}{\partial t}\right)^{2}-V(x)+x{\cal E}(t)+E\right]
\end{equation}
The contour of integration is shown in Fig.~7(b) where it starts at the point $t=0$. The horizontal dashed line 
corresponds to the conventional case ${\cal E}(t)=0$. The classical trajectory satisfies the equation in the complex time
\begin{equation}
\label{51}
m\hspace{0.1cm}\frac{\partial^{2}x}{\partial t^{2}}+V'(x)={\cal E}(t)
\end{equation}
For a smooth potential let us choose the signal in the form
\begin{equation}
\label{52}
{\cal E}(t)=-{\cal E}\hspace{0.1cm}\frac{t\theta}{t^{2}+\theta^{2}}\hspace{0.1cm}\exp\left(-\Omega^{2}t^{2}\right)
\end{equation}
We consider the case of the instant signal when $\Omega^{2}\theta^{2}\gg 1$ and ${\cal E}(t)$ is localized at the small 
part (2) of the order of $(\Omega^{2}\theta)^{-1}$ in Fig.~7(b). For the instant signal (\ref{52}) the motion along the 
part (1) in Fig.~7(b) (the horizontal path in Fig.~7(a)) is free and $x(t)=\tilde{x}(t)$, where the free trajectory 
$\tilde{x}(t)$ satisfies the equation
\begin{equation}
\label{53}
\frac{m}{2}\left(\frac{\partial\tilde{x}}{\partial t}\right)^{2}+V(\tilde{x})=E+\delta E\hspace{1cm}({\rm region}\hspace{0.1cm}1)
\end{equation}
A deviation of $m\partial x/\partial t$ from $m\partial\tilde{x}/\partial t$ starts at the region (2) in Fig.~7(b). If $t$ belongs to
the regions (2) (the vertical path in Fig.~7(a)) or (3) (the left horizontal path in Fig.~7(a)), one can write
\begin{equation}
\label{54}
m\hspace{0.1cm}\frac{\partial x}{\partial t}\simeq\int^{t}_{0}dt_{1}{\cal E}(t_{1})+
m\left(\frac{\partial\tilde{x}}{\partial t}\right)_{i\theta}\hspace{1cm}({\rm region}\hspace{0.1cm}2)
\end{equation}
The classical momentum $m\partial x/\partial t$ gets the jump $\int_{C}dt{\cal E}(t)$ after passing the part (2), but the 
coordinate $x$ does not have a jump on the time scale $(\Omega^{2}\theta)^{-1}$ in the vicinity of the part (2). At the 
region (3) the motion again is free, $x(t)=\tilde{x}(t)$, where the trajectory $\tilde{x}(t)$ obeys the equation
\begin{equation}
\label{55}
\frac{m}{2}\left(\frac{\partial\tilde{x}}{\partial t}\right)^{2}+V(\tilde{x})=E\hspace{1cm}({\rm region}\hspace{0.1cm}3)
\end{equation}
and has the form $\tilde{x}(t-i\tau_{0})$ (we omit a real time shift). Since, in the limit of an instant signal 
$\Omega^{2}\theta^{2}\gg 1$, the function $x(t)$ is almost continuous it follows that $\tau_{0}\simeq\theta$. Otherwise,
the real $x(t)$ would jump going from (1) to (3).

The classical energy gain
\begin{equation}
\label{56}
\delta E=-\int_{C}dt{\cal E}(t)\frac{\partial x}{\partial t}
\end{equation}
can be calculated by inserting Eq.~(\ref{54}) into Eq.~(\ref{56}). The result is
\begin{equation}
\label{57}
\delta E=-\left(\frac{\partial\tilde{x}}{\partial t}\right)_{i\theta}\int_{C}dt{\cal E}(t)-\frac{1}{2m}\left[\int_{C}dt{\cal E}(t)\right]^{2}
\end{equation}
Since, close to the point $i\theta$, $\exp(-\Omega^{2}t^{2})\simeq\exp[\Omega^{2}\theta^{2}-2i\Omega^{2}\theta(t-i\theta)]$
one can circle the contour $C$ around the point $t=i\theta$, which produces
\begin{equation}   
\label{58}
\delta E=\left(\pi\theta{\cal E}e^{\Omega^{2}\theta^{2}}\right)\left(\frac{\partial\tilde{x}}{\partial\tau}\right)_{\theta}+
\frac{1}{2m}\left(\pi\theta{\cal E}e^{\Omega^{2}\theta^{2}}\right)^{2}
\end{equation}
The velocity $\partial\tilde{x}/\partial\tau$ at $\tau =\theta$ is real and negative.

The duration (in the imaginary time) $\theta$ of the part (1) in Fig.~7(b) can be expressed through $\delta E$ according
to the classical formula
\begin{equation}
\label{59}
\theta =\sqrt{\frac{m}{2}}\int^{x_{1}}_{x_{0}}\frac{dx}{\sqrt{V(x)-E-\delta E}}
\end{equation}
Eq.~(\ref{59}), together with the boundary conditions $V(x_{1})=E+\delta E$ and $V(x_{0})=E$, determines the function 
$\delta E(E,\theta)$. At $t=i\theta$ the coordinate $x\simeq x_{0}$ and Eq.~(\ref{53}) results in
\begin {equation}
\label{60}
\delta E(E,\theta)=-\frac{m}{2}\left(\frac{\partial\tilde{x}}{\partial\tau}\right)^{2}_{\theta}
\end{equation}
It is remarkable that the value (\ref{60}) coincides with the extreme of $\delta E$ in Eq.~(\ref{58}) reached at
${\cal E}={\cal E}_{c}(E,\theta)$, where 
\begin{equation}
\label{61}
\pi\theta{\cal E}_{ext}(E,\theta)\exp\left(\Omega^{2}\theta^{2}\right)=\sqrt{-2m\delta E(E,\theta)}
\end{equation}
The action has the form
\begin{equation}
\label{62}
A=2\sqrt{2}\int^{x_{1}}_{x_{0}}dx\sqrt{V(x)-E-\delta E(E,\theta)}+2\theta\delta E(E,\theta)\hspace{0.1cm};\hspace{1cm}
\left({\cal E}={\cal E}_{ext}(E,\theta)\right)
\end{equation}
where the function $\delta E(E,\theta)$ is determined by Eq.~(\ref{59}). The condition $A=0$ defines the resonance energy 
$E_{R}(\theta)$ and Eq.~(\ref{61}) gives the threshold amplitude 
${\cal E}_{T}(\theta)={\cal E}_{ext}(E_{R}(\theta),\theta)$ 
\begin{equation}
\label{63}
{\cal E}_{T}(\theta)=\frac{\exp\left(\Omega^{2}\theta^{2}\right)}{\pi\theta}\sqrt{-2m\delta E(E_{R}(\theta),\theta)}
\end{equation}
If the signal amplitude is fixed, the expression (\ref{62}) is valid only for one value of energy $E=E_{ext}$ which is 
determined by the equation ${\cal E}={\cal E}_{ext}(E,\theta)$. If $E\neq E_{ext}$ the method of trajectories cannot be 
used and one should solve the Hamilton-Jacobi equation. ${\cal E}_{ext}$ is analogous to the same parameter (\ref{42}) and 
corresponds to the peak in the escape probability $W$ as in Fig.~5. One should note that whereas for a smooth potential 
the method of trajectories enables to obtain the result solely for one extreme energy, in the degenerated case of the 
singular potential (\ref{1}) this method is applicable in an interval of energies. This is due to that for the singular 
potential (\ref{1}) the maximum of $W(E)$ is sharp and the analog of the dashed curve of Fig.~8 coincides with the solid 
curve in Fig.~5. This does not mean the method of trajectories to be useless for a smooth potential since the energy 
$E_{ext}$ corresponds to a maximum of the escape rate and Eq.~(\ref{62}) determines the dashed line in Fig.~8. In the case 
of a smooth potential it is impossible to obtain an analytical solution of the Hamilton-Jacobi equation to get the 
preexponential factor in the expression for the decay rate $W$ as in the formula (\ref{18}). 
\section{SMOOTH POTENTIAL WELL IN AN ELECTRIC FIELD}
Eqs.~(\ref{59}) and (\ref{62}) allow to solve the problem of tunneling under action of an instant signal for a general
semiclassical potential $V(x)$. Let us consider a particular case of a smooth potential well in a weak electric field 
${\cal E}_{0}$. This means that in Fig.~7(a) $x_{0}\ll x_{1}$. In this situation one can use the approximation of 
triangular potential in Eq.~(\ref{59}) and (\ref{62})
\begin{equation}
\label{64}
A=\frac{4\theta}{3}\left[V-E-\delta E(E,\theta)\right]+2\theta\delta E(E,\theta)\hspace{0.1cm};\hspace{1cm}
\theta=\frac{\sqrt{2m(V-E-\delta E)}}{{\cal E}_{0}}
\end{equation}
The extreme electric field (\ref{61}) is given by the formula
\begin{equation}
\label{65}
\frac{\pi}{{\cal E}_{0}}{\cal E}_{ext}(E,\theta)\exp\left(\Omega^{2}\theta^{2}\right)=\sqrt{1-\frac{2m(V-E)}{\theta^{2}{\cal E}^{2}_{0}}}
\end{equation}
and the action (\ref{62}) has the form (\ref{41}) with the definition of $E_{R}$ (\ref{43})
\begin{equation}
\label{66}
A=2\theta\left[(V-E)-\frac{\theta^{2}{\cal E}^{2}_{0}}{6m}\right]\hspace{0.1cm};\hspace{1cm}({\cal E}={\cal E}_{ext}(E,\theta))
\end{equation}
The extreme value $E_{ext}$ in Fig.~8 is
\begin{equation}
\label{67}
E_{ext}=E_{R}-\frac{\theta^{2}{\cal E}^{2}_{0}}{2m}\left[\frac{2}{3}-
\left(\frac{\pi{\cal E}}{{\cal E}_{0}}\hspace{0.1cm}e^{\Omega^{2}\theta^{2}}\right)^{2}\right]\Theta\left({\cal E}_{T}-{\cal E}\right)
\end{equation}
where $E_{R}$ is given by Eq.~(\ref{43}) and the threshold value of the signal amplitude is
\begin{equation}
\label{68}
\frac{\pi{\cal E}_{T}}{{\cal E}_{0}}\hspace{0.1cm}\exp\left(\Omega^{2}\theta^{2}\right)=\sqrt{\frac{2}{3}}
\end{equation}
If the energy level $E$ in the well is fixed, the decay rate $W$ is determined by Eq.~(\ref{48}) with the resonance value
of the signal parameter $\theta_{R}$ given by Eq.~(\ref{47}).

The condition of applicability of the semiclassical approximation cannot be derived for a smooth potential with the all 
details, as the condition (\ref{28}), since for this purpose one should know the solution of the Hamilton-Jacobi equation.
Nevertheless, the typical feature of that condition $(V-E)t\gg 1$, where $t$ is a characteristic time of the problem, can 
be used as an approximation. Since the typical time is $(\Omega^{2}\theta)^{-1}$, the semiclassical condition reads
\begin{equation}
\label{69}
\Omega^{2}\theta^{2}\ll (V-E)\theta
\end{equation}
The condition (\ref{69}) does not contradict to a big value of $\Omega^{2}\theta^{2}$ since in the semiclassical case 
$(V-E)\theta$ is a big parameter.
\section{OVER-BARRIER ESCAPE}
A metastable state in a well can decay also by the over-barrier transition picking up a number of quanta necessary to 
reach the top of the barrier (in our case there is no a thermal activation). This process is competitive with tunneling 
under the barrier. Let us estimate the probability of the over-barrier escape. The Fourier component of the signal 
(\ref{52}) 
\begin{equation}
\label{70}
{\cal E}_{\omega}=-\theta{\cal E}\exp\left(-\frac{\omega^{2}}{4\Omega^{2}}\right)\int^{\infty}_{-\infty}\frac{tdt}{t^{2}+\theta^{2}}
\exp\left[-\Omega^{2}\left(t-\frac{i\omega}{2\Omega^{2}}\right)^{2}\right]
\end{equation}
is determined by the saddle contribution (the horizontal path ${\rm Im}t=\omega/2\Omega^{2}$) and by the pole integration 
at $t=i\theta$, if $\theta <\omega/2\Omega^{2}$
\begin{equation}
\label{71}
{\cal E}_{\omega}=-i\pi\theta{\cal E}\exp\left(\Omega^{2}\theta^{2}-\omega\theta\right)\Theta\left(\frac{\omega}{2\Omega^{2}}-\theta\right)
+2i\sqrt{\pi}\frac{{\cal E}\omega\theta\Omega}{\omega^{2}-(2\Omega^{2}\theta)^{2}}\exp\left(-\frac{\omega^{2}}{4\Omega^{2}}\right)
\end{equation}
The probability of the over-barrier escape $W_{o-b}$ is proportional to the absorption of $N$ quanta which can be 
estimated as $({\cal E}_{\omega})^{2N}$ and can be written schematically in the form
\begin{equation}
\label{72}
W_{o-b}\sim{\cal E}^{2N}\left\{\exp\left[2N(\Omega^{2}\theta^{2}-\omega\theta)\right]+\exp\left(-2N\frac{\omega^{2}}{4\Omega^{2}}\right)\right\}
\end{equation}
where $N=(V-E)/\omega$ and we ignore an interference of the two contributions. Since the typical signal amplitude is
${\cal E}\sim\exp(-\Omega^{2}\theta^{2})$, the probability (\ref{72}) reads
\begin{equation}
\label{73}
W_{o-b}\sim\exp\left[-2\theta(V-E)\right]+
\exp\left[-2(V-E)\left(\frac{\omega^{2}}{4\Omega^{2}}+\frac{\Omega^{2}\theta^{2}}{\omega}\right)\right]
\end{equation}
The optimum frequency $\omega =2\Omega^{2}\theta$ in the second term produces the same exponent as the first one. 
Therefore, the probability of the over-barrier escape, stimulated by the signal (\ref{52}), can be estimated as 
\begin{equation}
\label{74}
W_{o-b}\sim\exp\left[-2\theta(V-E)\right]
\end{equation}
The exponent in the expression (\ref{74}) is of the same magnitude as the action $A$ far away from the resonance. For this 
reason, an escape mechanism under the action of the signal (\ref{52}) is the tunneling process but not an over-barrier 
escape. From the mathematical point of view, it is understandable since the extreme of the action $A$ is provided by the 
negative energy gain $\delta E$ but not by $\delta E=(V-E)$. 
\section{MANIFESTATIONS OF THE EUCLIDEAN RESONANCE}
In this section we point out some particular physical situations when specially adapted signals result or may result in a
dramatically enhanced tunneling rate due to the Euclidean resonance.
\subsection{Selective ionization of atoms}
As a simplest example of the Euclidean resonance let us consider the ionization of the hydrogen atom by the signal 
(\ref{52}). In this case one can use Fig.~7(a), considering the well to be extended in energy down to $-\infty$. Suppose the
constant electric field ${\cal E}_{0}=2\times {10}^{7}{\rm eV}/{\rm cm}$. In this field the conventional tunneling rate is 
proportional to $\exp\left(-A_{0}\right)=\exp(-76)$, where $A_{0}$ is given by Eq.~(\ref{20}) and 
$(V-E)\simeq13.6{\rm eV}$. This provides the decay time $(V-E)^{-1}\exp\left(76\right)\sim{10}^{9}{\rm years}$. Under the action of the 
signal with the amplitude ${\cal E}>{\cal E}_{T}$, where the threshold field is given by Eq.~(\ref{68}), the Euclidean 
resonance occurs at 
\begin{equation}
\label{75}
\theta_{R}=\frac{\sqrt{6m(V-E)}}{{\cal E}_{0}}\simeq{10}^{-14}{\rm s}
\end{equation}
and the tunneling rate is determined by Eq.~(\ref{48}) which reads in this case
\begin{equation}
\label{76}
W\sim{10}^{-170(\theta_{R}-\theta)/\theta_{R}}\Theta\left(\theta_{R}-\theta\right)
\end{equation}
According to Eq.~(\ref{45}), $(V-E-\delta E)=3(V-E)$. With the parameters chosen, $x_{1}/x_{0}\simeq 100$ and the duration 
of the exit wave packet (\ref{46}) is $\delta t\simeq 0.28\theta\simeq 2.8\times{10}^{-15}{\rm s}$. Using the semiclassical
condition (\ref{69}), which now reads $\Omega^{2}\theta^{2}\ll 200$, one can take the signal parameter 
$\Omega^{2}\theta^{2}=15$. This gives ${\cal E}_{T}\simeq 1.5 {\rm eV}/{\rm cm}$ and 
$\Omega^{-1}\simeq 2.6\times{10}^{-15}{\rm s}$. For a smaller constant electric field ${\cal E}_{0}$, the resonance value 
$\theta_{R}$ and the threshold signal amplitude ${\cal E}_{T}$ are also smaller. As follows from the relation (\ref{76}), 
the uncertainty of the signal parameter $\theta$ should be approximately $1\%$ to be close to the resonance. When in the 
system there are a few types of atoms/molecules, one can make a selective ionization of a particular group of them.
\subsection{Emission of electrons from a metallic surface}  
Electrons can tunnel from a metallic surface when an electric field ${\cal E}_{0}$ is applied perpendicular to it. In this 
case $V$ is the work function and $E$ can be at the Fermi energy or down. When the constant electric field is modulated by
a successive set of signals (\ref{52}) separated by the time interval of the order of $\theta$, there is the steady current
of electrons distributed around the energy $E+\delta E$ with the relative accuracy of approximately $A^{-1}_{0}$. The 
current density can be estimated as $j\sim env_{F}/A_{0}$, where $n$ is the electron density in the metal and $v_{F}$ is 
the Fermi velocity. For the parameters of Sec. XI(A), the steady electric current from the metal surface is
$j\sim {10}^{8}{\rm A}/{\rm cm}^{2}$. If to use Eq.~(\ref{20}) for $A_{0}$ the current density can be written in the form
$j=\sigma {\cal E}_{0}/e$, where the conductivity is $\sigma\sim e^{2}n/m(V-E)$. When the metallic sample is connected to a
circuit, its electroneutrality recovers faster comparing to the time (\ref{75}). 
\subsection{Selective stimulation of chemical reactions}
A variety of chemical and biochemical reaction are characterized by the potential barrier for electrons which separates the
electron state of the energy $E$ before reaction and the state of the energy $E_{1}$ after reaction $(E_{1}<E)$ 
\cite{VOLK}. Usually, the tunneling rate for chemical reactions is negligible and they occur by thermal activation through 
the barrier which requires, in many cases, a reduction by a catalyst. Tunneling, stimulated by the Euclidean resonance, 
can provide new possibilities for reactions which are impossible under conventional conditions. Under the conditions of 
the resonance, the exit energy $E+\delta E$ does not necessary coincides with $E_{1}$ but during the motion of molecules 
towards each other the energy parameters vary and at some distance the condition $E+\delta E = E_{1}$ can be encountered. 
Using the Euclidean resonance, one can select some particular bonds by the resonance to choose a type of reaction to be 
stimulated.

By this method one can control also some nonvalent molecular conversions such as internal rotations of a part of the
molecule around a single bond, etc. In this case the Euclidean resonance stimulates tunneling of ions. For ionic tunneling
the resonance $\theta_{R}$ is approximately two orders of magnitude bigger and ${\cal E}_{T}$ is smaller comparing to 
electron tunneling.
\subsection{Stimulation of alpha-decay}
In the problem of alpha-decay the potential energy of the alpha-particle and the nucleus is shown in Fig.~9, where the
potential to the right of the nucleus has the form $\beta/x$ and the typical nucleus size is 
$x_{0}\sim{10}^{-13}{\rm cm}$. The conventional decay rate is $W\sim\exp(-A_{0})$, where
\begin{equation}
\label{77}
A_{0}=\pi\beta\sqrt{\frac{2m}{E}}
\end{equation}
and $m$ is the mass of the alpha-particle \cite{LANDQUANT}. A properly adapted resonance signal can result in huge 
enhancement of alpha-decay. In a situation of the Euclidean resonance Eq.~(\ref{59}) reads (with 
$x_{0}\simeq 0$)
\begin{equation}
\label{78}
\theta = \frac{\pi}{4}\hspace{0.1cm}\frac{\beta\sqrt{2m}}{(E+\delta E)^{3/2}}
\end{equation}
The action (\ref{62}) has the form
\begin{equation}
\label{79}
A=\left(3\pi\beta\sqrt{3m\theta}\right)^{2/3}-2\theta E
\end{equation}
At the Euclidean resonance $(A=0)$, 
\begin{equation}
\label{80}
\theta_{R}=\pi\beta\sqrt{\frac{27m}{8E^{3}}}\hspace{0.1cm};\hspace{1cm}\frac{E+\delta E}{E}=\frac{1}{3}
\end{equation}
where the energy $E+\delta E$ corresponds to the horizontal path in Fig.~9. The decay rate, within the exponential 
accuracy, is
\begin{equation}
\label{81}
W\sim\exp\left(-\sqrt{3}A_{0}\frac{\theta_{R}-\theta}{\theta_{R}}\right)\Theta\left(\theta_{R}-\theta\right)
\end{equation}
The threshold amplitude of the signal (\ref{63}) is
\begin{equation}
\label{82}
{\cal E}_{T}=\frac{4\sqrt{2}}{9\pi^{2}}\hspace{0.1cm}\frac{E^{2}}{\beta}\exp\left(-\Omega^{2}\theta^{2}\right)
\end{equation}
For the decay ${\rm Nd}^{144}_{60}\rightarrow{\rm Ce}^{140}_{58}+\alpha$, one should take $\beta =2\times 58 e^{2}$ and 
$E=1.9{\rm MeV}$. This gives $\theta_{R}\simeq 0.8\times{10}^{-19}{\rm s}$, $A_{0}\simeq177$, and (with 
$\Omega^{2}\theta^{2}=15$) ${\cal E}_{T}\simeq{10}^{9}{\rm eV}/{\rm cm}$. These parameters unlikely are suitable for an 
experiment. Nevertheless, the situation may be different. The energy of an alpha-particle $E-E_{0}$ is always in the 
Mev-scale and the experimentally observed $E$ varies approximately between 2 and 10 Mev for different nuclei. This does 
not exclude a situation when for some nuclei the energy of the alpha-particle $E$ can be of the order of 1KeV. Normally, 
alpha-decay of such nuclei hardly can be detected in a conventional way due to an extremely small decay rate since 
$A_{0}\simeq 1.3\times{10}^{4}$ (\ref{77}). But under the conditions of the Euclidean resonance for $E=1{\rm Kev}$, the 
resonance value is $\theta_{R}\simeq{10}^{-14}{\rm s}$ (\ref{78}) and one can obtain the flux of alpha-particles with the energy 
$1/3{\rm KeV}$ and the approximate duration ${10}^{-15}{\rm s}$ per one signal (\ref{46}). One can take 
$\Omega^{2}\theta^{2}=18$ which gives ${\cal E}_{T}\simeq 20{\rm eV}/{\rm cm}$ (\ref{81}). The relative accuracy of the signal 
parameter $\theta$ near the resonance, according to Eq.~(\ref{81}), should be $0.01\%$ but $\theta$, of the order of ten 
femtoseconds, relates with experimental facilities. In this way, one can search nuclei exhibiting the soft alpha-decay.
\section{SIGNAL CONDITIONS}
The signal shape (\ref{52}) is not unique, nevertheless, we focus on this shape. The plot at $\Omega^{2}\theta^{2}=15$ is 
shown in Fig.~10 where the vertical and the horizontal axes correspond to ${\cal E}(t)/{\cal E}$ and $t/\theta$. A 
deviation of the signal from the dependence
\begin{equation}
\label{83}
\tilde{\cal E}(t)=-{\cal E}\frac{t}{\theta}\exp\left(-\Omega^{2}t^{2}\right)
\end{equation}
is less than $10\%$, however, the signal (\ref{83}) does not lead to a Euclidean resonance, resulting only in a 
perturbative correction to the conventional formula $W\sim\exp(-A_{0})$. From mathematical point of view, it is due to 
that the signal (\ref{83}) does not have a pole at a complex time. We do not study here in full how fluctuations of the 
signal influence the resonance. For the resonance condition it should be 
$(\theta_{R}-\theta)/\theta_{R}\lesssim A^{-1}_{0}$ (\ref{48}) which determines the signal accuracy with respect to the 
parameter $\theta$.

The amplitude of the threshold signal ${\cal E}_{T}$ is proportional to $\exp\left(-\Omega^{2}\theta^{2}\right)$. Due to 
the semiclassical condition (\ref{69}), the parameter $\Omega^{2}\theta^{2}\ll A^{-1}_{0}$ can be big for very 
non-transparent (without a non-stationary signal) barriers, leading to a very small threshold signal. In the limit of a 
non-transparent barrier $(A_{0}=\infty)$, the amplitude of the threshold signal is ${\cal E}=0$. What is a restriction for 
this non-physical limit?

Every physical system, strictly speaking, is not Hamiltonian and the energy levels always have a finite width due to
interaction with other subsystems. In the classical limit the equation of motion contains the friction coefficient 
$\gamma$ 
\begin{equation}
\label{84}
m\hspace{0.1cm}\frac{\partial^{2}x}{\partial t^{2}}+m\gamma\hspace{0.1cm}\frac{\partial x}{\partial t}+V'(x)=0
\end{equation}
A physical origin of $\gamma$ depends on a particular situation. According to the theory of dissipative quantum mechanics 
of Caldeira and Leggett \cite{CALDLEGG} (see also Ref.~\cite{HANGGI}), a friction disturbs the under-barrier motion 
resulting in the energy-loosing path (1a) shown in Fig.~7(a). The frictionless path (1) and the friction path (1a) 
correspond to the same time $\theta$ and, due to this, the energy gain $\mid\delta E\mid$ is less for the path (1a). This 
makes impossible the resonance condition $(A=0)$ when the friction becomes sufficiently big $\gamma >\gamma_{c}$. The 
approximate estimate of the critical friction is $\gamma_{c}\sim {\theta}^{-1}$. Therefore, the Euclidean resonance exists
 for a dissipative system if the friction coefficient is not big
\begin{equation}
\label{85}
\gamma <\gamma_{c}\sim{\theta}^{-1}
\end{equation}
The threshold amplitude can be estimated in a general case as 
${\cal E}_{T}\sim {\cal E}_{0}\exp\left(-\Omega^{2}\theta^{2}\right)$ where ${\cal E}_{0}$ is some typical electric field 
in the system. The restriction condition is
\begin{equation}
\label{86}
\Omega^{2}\theta^{2}\ll(V-E)\theta\lesssim\frac{V-E}{\gamma}
\end{equation}
The main feature of the signal (\ref{52}) is its Lorentzian shape which results in the poles at the complex time. The 
Gaussian modulation is not obligatory and results only to the strong reduction of the threshold amplitude ${\cal E}_{T}$.
We do not consider here technical possibilities to tailor proper electromagnetic signals which should be in the 
approximate range of 1-100 fs. Recent achievements in formation of short signals are presented in 
Ref.~\cite{WEINER,EFIM,KAPLAN1,HARRIS,PAPA,HERTZ,KAPLAN2}.  
\section{DISCUSSIONS AND CONCLUSIONS}
The probability of tunneling through an almost classical static barrier is exponentially small, according to the general
principles of quantum mechanics. Nevertheless, a particle can penetrate through a very non-transparent barrier under the
action of the specially adapted signal of an external field. The signal parameters select the particle energy $E_{R}$. 
For particle energies close to this resonance value, the tunneling rate is not small during some finite interval of time 
and has a very sharp peak at the resonance energy $E_{R}$.  The width of the peak $(E_{R}-E)/E_{R}\sim A^{-1}_{0}(E_{R})$ 
is connected to the static action $A_{0}(E_{R})$ which can be of the order of hundred or even bigger for a very 
non-transparent barrier. After entering inside the barrier the particle emits electromagnetic quanta and exit the barrier 
with a lower energy. It is reasonable to call this phenomenon the Euclidean resonance since the motion occurs in imaginary
(complex) time. 

The Euclidean resonance can have various applications to the problems where quantum tunneling is an essential process. For
example, chemical and biochemical reactions occur by thermal activation over a potential barrier which, in many cases, 
should be lowered by a catalyst. The potential chemical conversions, which cannot be realized due to high barriers, may 
get run by the effect of the Euclidean resonance. Since the resonance peak is very sharp, one can use the effect for 
selective stimulation of chemical reactions adapting the signal to some particular chemical bond(s). Another possible 
application of the Euclidean resonance is search of the soft alpha-decay when emitting alpha-particles have the energy in
the interval of 0.1-10 KeV. The duration of the required signal is in the range 1-10 fs and the amplitude is of the order
of 1-10 eV/cm. The soft alpha-decay, if it exists, cannot be observed in the conventional way due to extremely small decay
rate.

The phenomenon of the Euclidean resonance may occur not only in tunneling processes but also in the over-barrier 
reflection of particles in quantum mechanics \cite{LANDQUANT} and in classical reflection of macroscopic waves 
(electromagnetic, acoustic, etc.) from a spatially-smooth medium \cite{HEADING}. All these effects constitute a class of the
Stokes phenomena since in a static case they have the common mathematic feature: the classical turning point generates the
Stokes lines \cite{HEADING} where a jump of the sub-dominant (exponentially small) solution occurs.

\acknowledgments
I am grateful to J. Engelfried for providing the experimental data on alpha-decay.

\newpage
\begin{figure}[p]
\begin{center}
\vspace{1.5cm}
\epsfxsize=\hsize
\epsfxsize=10cm
\leavevmode
\epsfbox{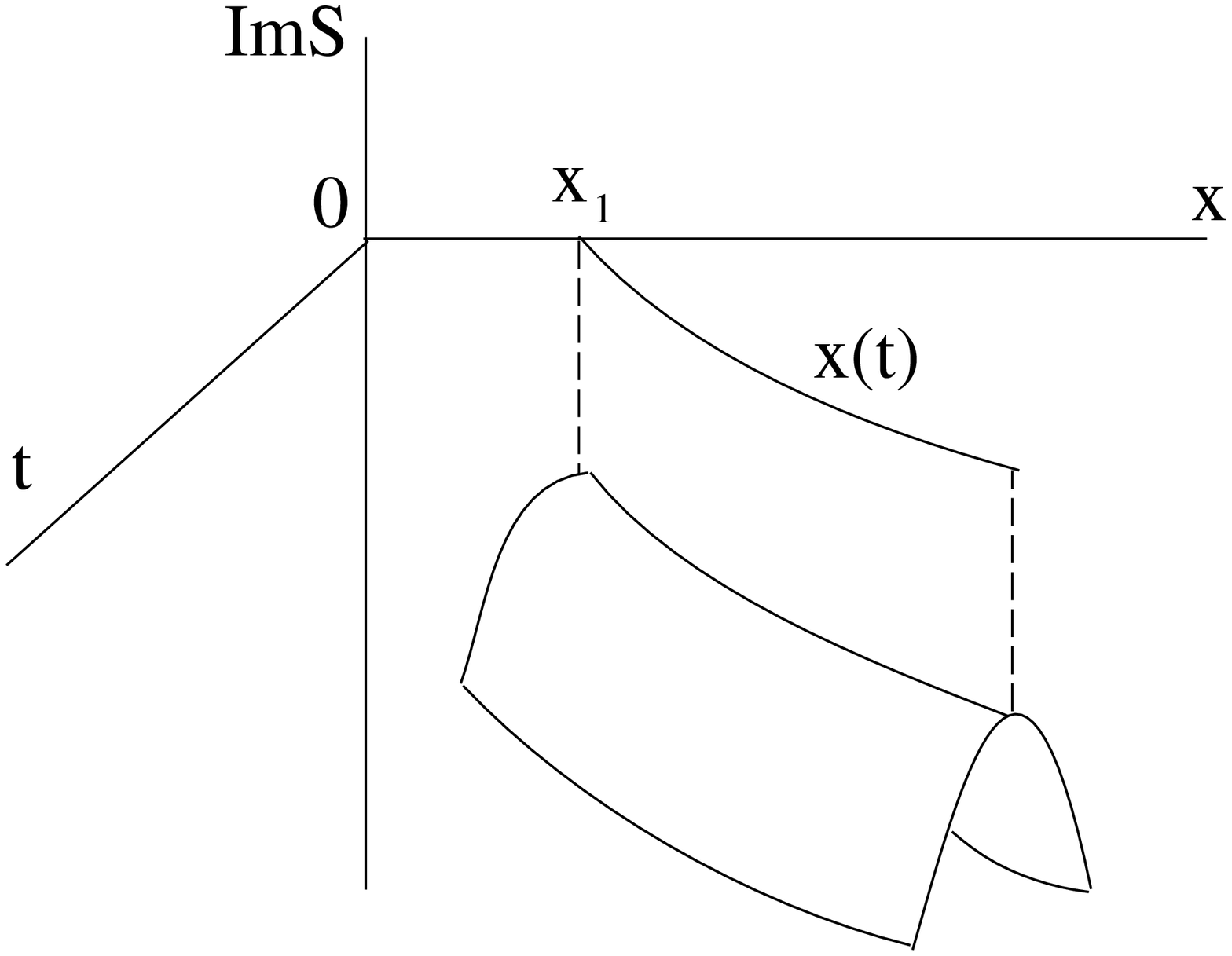}
\vspace{2cm}
\caption{The imaginary part of the action is a constant at the real classical trajectory $x(t)$ after exit from the barrier.
Only a part of the $\{x;t\}$-plane in the vicinity of the trajectory is shown.}
\label{fig1}
\end{center}
\end{figure}

\newpage
\begin{figure}[p]
\begin{center}
\vspace{3cm}
\epsfxsize=\hsize
\epsfxsize=14cm
\leavevmode
\epsfbox{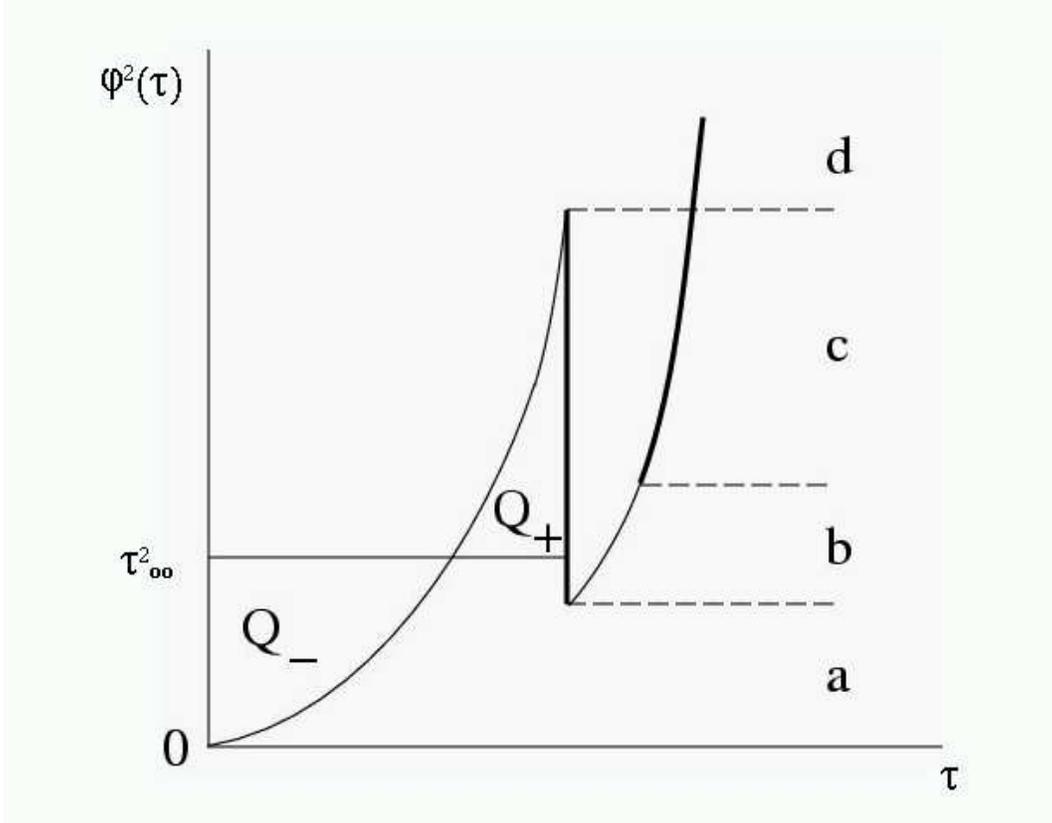}
\vspace{1.5cm}
\caption{The plot of the function $\varphi^{2}(\tau)$ consists of the two parabolas $\tau^{2}$ and 
$(\tau -\lambda\theta)^{2}$ connected by the vertical line. The solution $\tau =\tau_{1}$ of the equation 
$\varphi^{2}(\tau)=\tau^{2}_{00}$ determines the exit point $x_{1}$  where $\partial S(x,0)/\partial x=0$. Only the thick 
curves are physical. The same graph gives the geometrical condition (the equality of the areas $Q_{+}=Q_{-}$) for the 
Euclidean resonance.}
\label{fig2}
\end{center}
\end{figure}

\newpage
\begin{figure}[p]
\begin{center}
\vspace{3cm}
\epsfxsize=\hsize
\epsfxsize=11cm
\leavevmode
\epsfbox{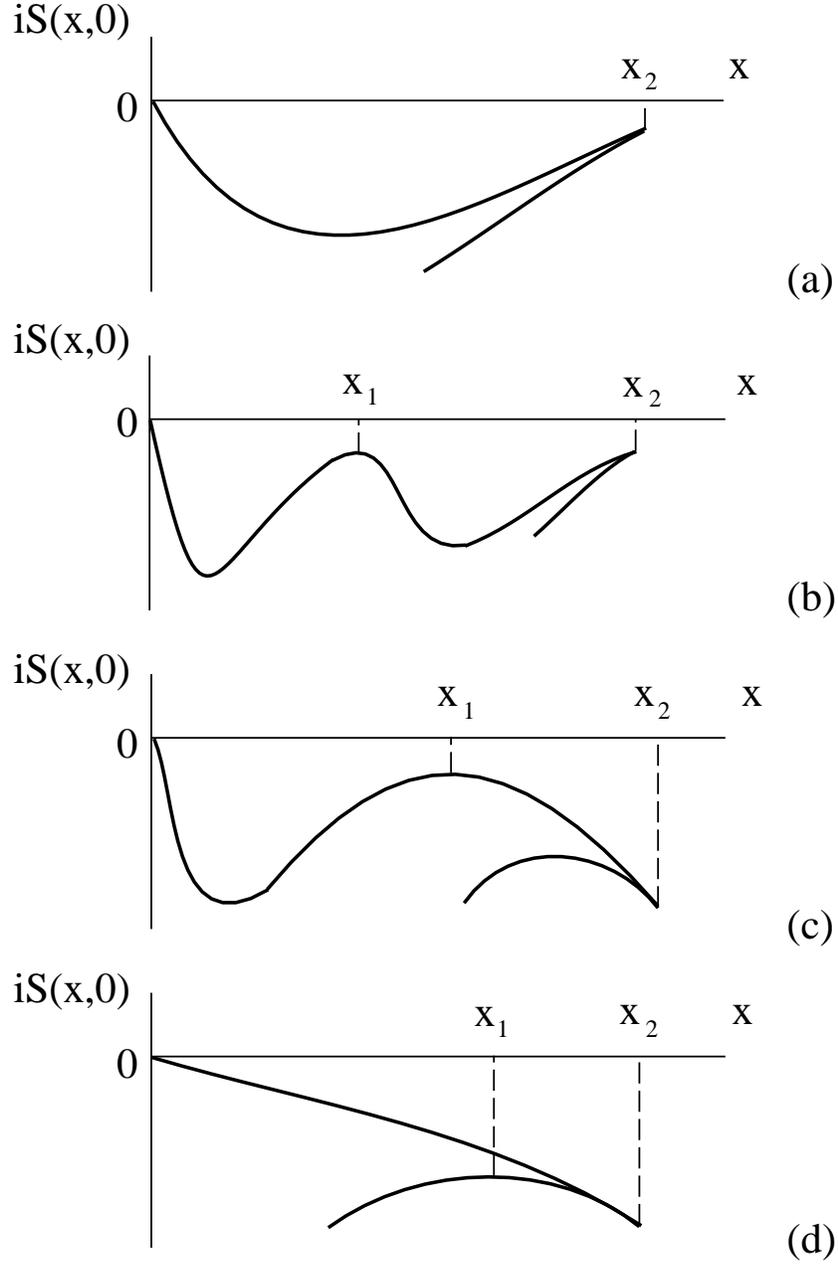}
\vspace{1.5cm}
\caption{The action for different domains of $\tau^{2}_{00}$ in Fig.~2. $x_{1}$ corresponds to the stable regime with
$i\partial^{2}S/\partial x^{2}<0$ denoted in Fig.~2 by the thick solid curves.}
\label{fig3}
\end{center}
\end{figure}

\newpage
\begin{figure}[p]
\begin{center}
\vspace{3cm}
\epsfxsize=\hsize
\epsfxsize=10cm
\leavevmode
\epsfbox{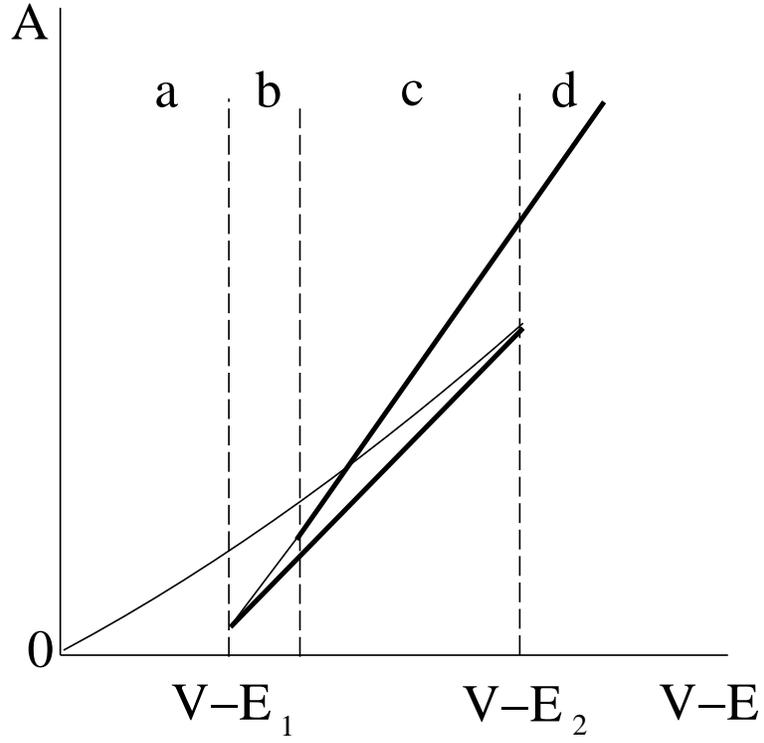}
\vspace{1.5cm}
\caption{The action for the potential (32). The thick solid curves correspond to the physical regimes.
$V-E_{1}=(V-E_{2})(1-\lambda)^{2}$ and $V-E_{2}=\theta^{2}{\cal E}^{2}_{0}/2m$.}
\label{fig4}
\end{center}
\end{figure}

\newpage
\begin{figure}[p]
\begin{center}
\vspace{3cm}
\epsfxsize=\hsize
\epsfxsize=13cm
\leavevmode
\epsfbox{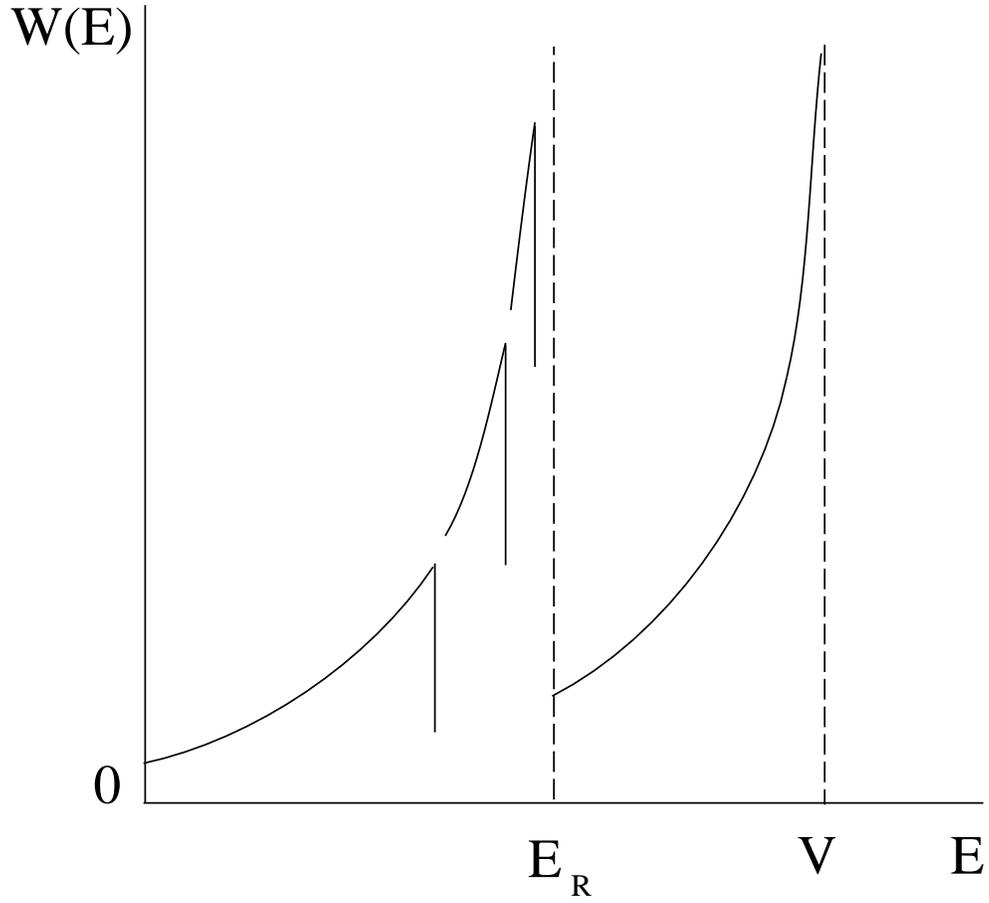}
\vspace{1.5cm}
\caption{At each amplitude of the signal, there is a peak of $W$ at the energy $E_{ext}$. When 
$\lambda\geq\lambda_{R}$, the peak reaches its maximum value at $E=E_{R}$. When $E>E_{R}$, $W$ is determined by the 
WKB formula.}
\label{fig5}
\end{center}
\end{figure}

\newpage
\begin{figure}[p]
\begin{center}
\vspace{3cm}
\epsfxsize=\hsize
\epsfxsize=10cm
\leavevmode
\epsfbox{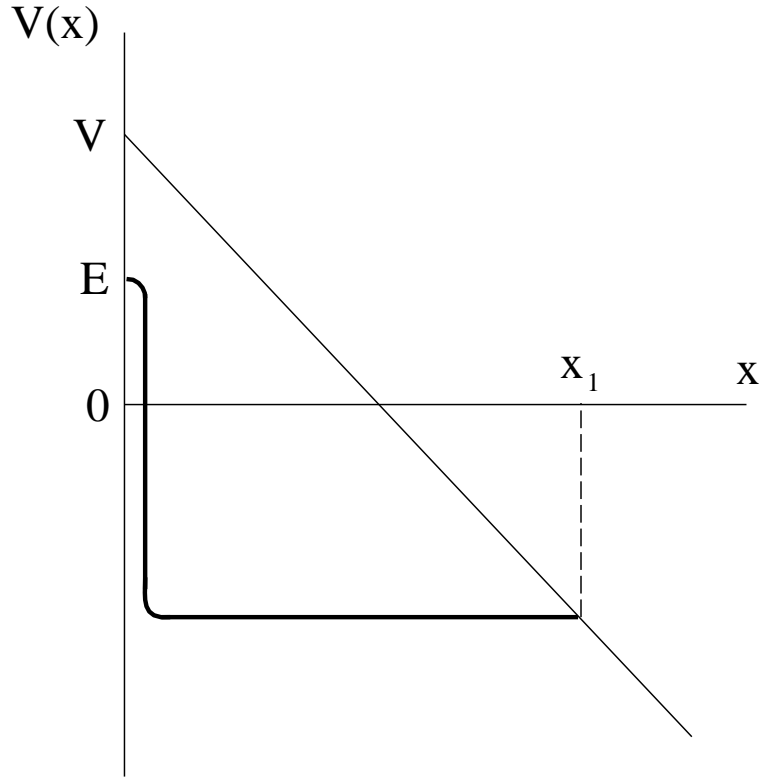}
\vspace{1.5cm}
\caption{The total energy of the particle under the barrier. The particle starts at the point $x_{1}$ ($\tau =0$), reaches 
the point $x\simeq 0$ at the ``moment'' $\tau\simeq\theta $, acquires instantly the energy gain, and reaches the point
$x=0$ having the energy $E$. At the horizontal path the particle energy is $E+\delta E$.}
\label{fig6}
\end{center}
\end{figure}

\newpage
\begin{figure}[p]
\begin{center}
\vspace{3cm}
\epsfxsize=\hsize
\epsfxsize=12cm
\leavevmode
\epsfbox{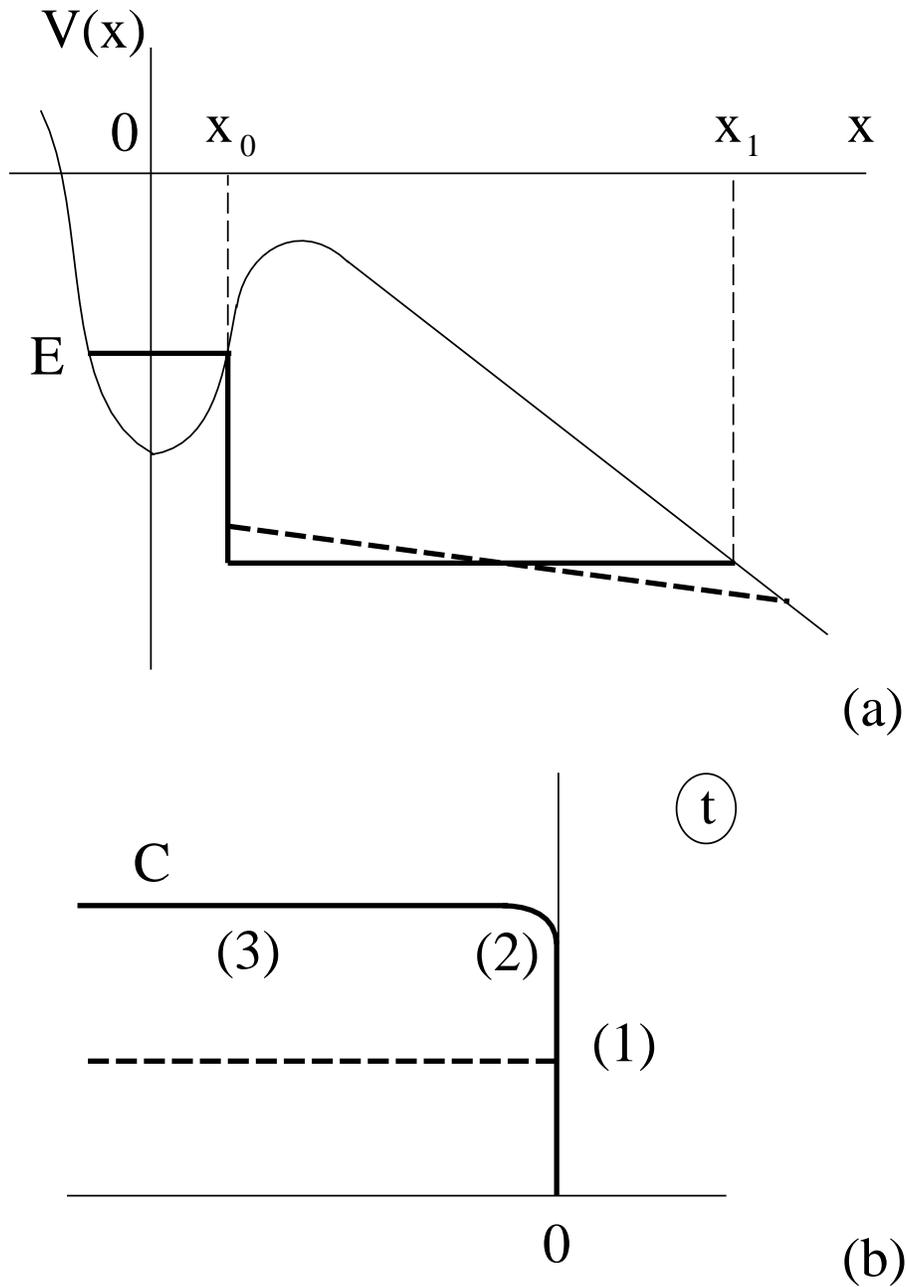}
\vspace{1.5cm}
\caption{(a) The path of the particle under the barrier in terms of its total energy. The signal acts at the ``diving'' 
(vertical) part. The dashed curve shows the particle energy in the damped case. (b) In terms of the complex time, the part
(1) represents the horizontal path in Fig.~7(b), the part (2) - the vertical path, and the part (3) - the left horizontal 
path.}
\label{fig7}
\end{center}
\end{figure}

\newpage
\begin{figure}[p]
\begin{center}
\vspace{3cm}
\epsfxsize=\hsize
\epsfxsize=15cm
\leavevmode
\epsfbox{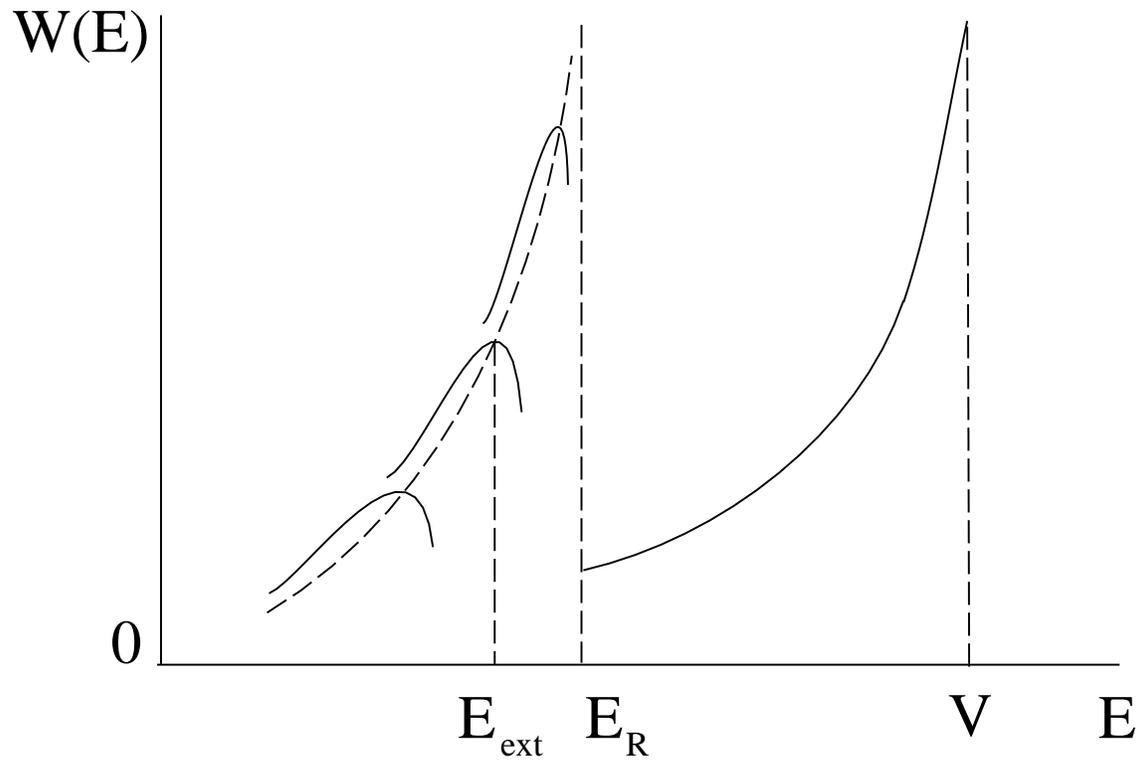}
\vspace{1.5cm}
\caption{The tunneling rate $W\sim\exp(-A)$ at each fixed amplitude of the signal ${\cal E}<{\cal E}_{T}$ is a curve with
the maximum at $E=E_{ext}$. $E_{ext}=E_{R}$ at ${\cal E}={\cal E}_{T}$. $W$ reaches its absolute maximum at $E=E_{R}$
under the condition ${\cal E}\geq{\cal E}_{T}$.}
\label{fig8}
\end{center}
\end{figure}

\newpage
\begin{figure}[p]
\begin{center}
\vspace{3cm}
\epsfxsize=\hsize
\epsfxsize=10cm
\leavevmode
\epsfbox{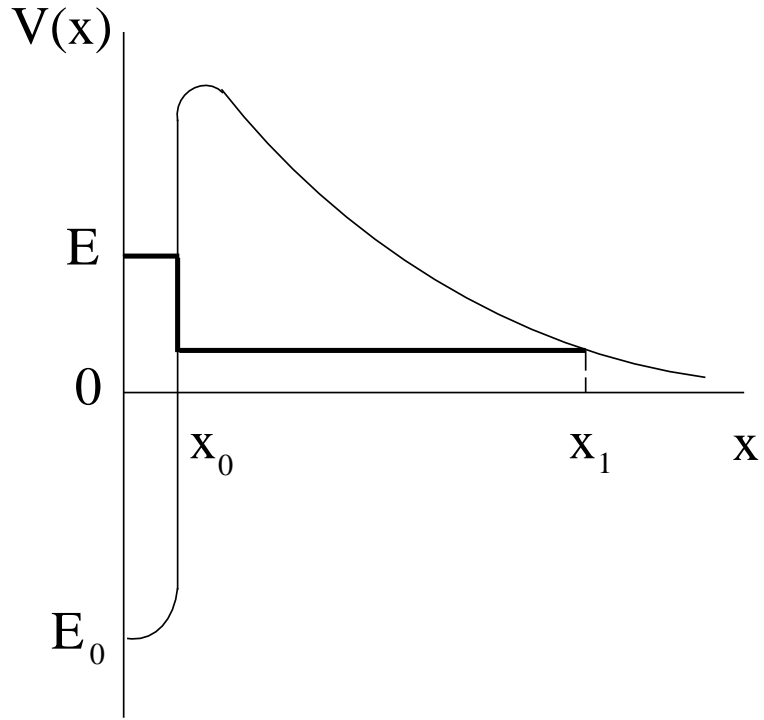}
\vspace{1.5cm}
\caption{The potential energy of the alpha-particle and the nucleus. Due to the Euclidean resonance, the energy of the exit
alpha-particle (the lower horizontal path) is $E+\delta E=E/3$.}
\label{fig9}
\end{center}
\end{figure}

\newpage
\begin{figure}[p]
\begin{center}
\vspace{3cm}
\epsfxsize=\hsize
\epsfxsize=10cm
\leavevmode
\epsfbox{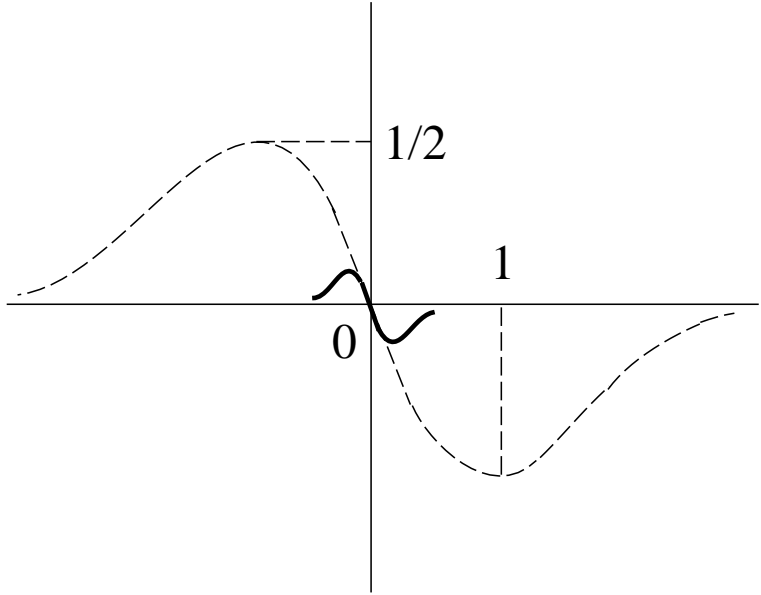}
\vspace{1.5cm}
\caption{The signal (52) is drawn by the solid curve at $\Omega^{2}\theta^{2}=15$. The vertical and the horizontal axes correspond
to ${\cal E}(t)/{\cal E}$ and $t/\theta$. The dashed curve shows the signal 
${\cal E}(t)/{\cal E}=-(t/\theta)\left[1+(t/\theta)^{2}\right]^{-2}$.}
\label{fig10}
\end{center}
\end{figure}

\end{document}